\documentclass[preprintnumbers,amsmath,amssymbm,prd]{revtex4}
\usepackage{epsfig}
\usepackage{graphicx}

\begin{document}
\title{The quantum mass gap of extremal black holes}
\author{Shahar Hod}
\address{The Ruppin Academic Center, Emeq Hefer 40250, Israel}
\address{ }
\address{The Hadassah Institute, Jerusalem 91010, Israel}
\date{\today}

\begin{abstract}
\ \ \ Using sophisticated string theory calculations, Maldacena and
Susskind have intriguingly shown that near-extremal black holes are characterized by
a {\it finite} mass gap above the corresponding zero-temperature (extremal) black-hole configuration.
In the present compact paper we explicitly prove that the minimum energy
gap ${\cal E}_{\text{gap}}=\hbar^2/M^3$, which
characterizes the mass spectra of near-extremal charged Reissner-Nordstr\"om black holes, can be
inferred from a simple semi-classical analysis.
\end{abstract}
\bigskip
\maketitle


\section{Introduction}

Extremal black holes are of fundamental importance in candidate
theories of quantum gravity since they mark the boundary between
black-hole configurations and horizonless naked
singularities \cite{Cohc,Hisc,Clarc,BekRosc,Bradc,Hub,Hod1c,Hod2c,Hod3c}.
These unique solutions of the Einstein field equations, which are
characterized by the minimally allowed mass (radius) for a given
amount of black-hole charge or angular momentum
\cite{Noteunits}, have attracted much attention from physicists and
mathematicians during the last five decades.

In particular, extremal black holes play a key role in various types
of gedanken experiments that have been designed to challenge the
Penrose cosmic censorship conjecture \cite{Pen,HawPen,Cohc,Hisc,Clarc,BekRosc,Bradc,Hub,Hod1c,Hod2c,Hod3c}.
Likewise, extremal black holes play a fundamental role in various attempts to prove the validity of the intriguing
weak gravity conjecture \cite{Arkwk,Chewk,Kamwk,Browk,Kapwk,Hebwk,Koowk,Klawk,Hodwk}.
In addition, the thermodynamic and statistical properties of extremal
black holes have been extensively studied by many physicists in
order to gain some insights on the origin of the Bekenstein-Hawking
entropy \cite{Strob,Calb,Horb,Hawnull,Teitnull,Ghonull,Hodnull}.

One of the most intriguing predictions of string theory is that near-extremal black holes are
characterized by a {\it finite} (non-zero) energy gap of the first excited state above the extremal (zero-temperature)
configuration \cite{Gap1}. In particular, it has been proved \cite{Gap2} that
charged Reissner-Nordstr\"om (RN) black holes are characterized by the finite mass gap \cite{Noteq0,Notemq}
\begin{equation}\label{Eq1}
{\cal E}_{\text{gap}}\equiv M-M_{\text{extremal}}={{\hbar^2}\over{M^3}}\ \ \ \ ; \ \ \ \ M_{\text{extremal}}=Q\
\end{equation}
of the first excited state above the extremal $M_{\text{extremal}}=Q$ black-hole configuration.

It is important to emphasize that the authors of the physically interesting work \cite{Gap2} have derived 
the suggested energy gap (\ref{Eq1}) using a physical argument which is based on holography 
without using explicit string theory calculations. In the present paper we shall argue 
that an energy gap of the form (\ref{Eq1}) arises naturally in a gedanken experiment that 
involves neither string theory calculations nor holography.

The main goal of the present compact paper is to gain some physical
insights on the origins of the physically intriguing mass gap
(\ref{Eq1}). In particular, below we shall demonstrate that the
interesting mass gap (\ref{Eq1}) can be inferred from a simple
semi-classical gedanken experiment [we would like to stress the fact that we do not rule out the 
possible existence of an energy gap for extremal black holes which 
may be smaller than the suggested gap (\ref{Eq1})]. 
To this end, we shall analyze
the physical and mathematical properties of composed
extremal-RN-black-hole-massive-shell configurations.

\section{Description of the system}

We consider a physical system which is composed of a charged RN
black hole of mass $M$ and electric charge $Q$ and a concentric spherically
symmetric massive neutral shell of radius $R$ and energy (energy as measured by asymptotic observers) $E(R)$.
According to the Birkhoff theorem, the spacetime inside the
spherically symmetric shell is described by the curved RN line
element \cite{Whb,Notesch}
\begin{equation}\label{Eq2}
ds^2=-\Big(1-{{2M}\over{r}}+{{Q^2}\over{r^2}}\Big)dt^2+\Big(1-{{2M}\over{r}}+{{Q^2}\over{r^2}}\Big)^{-1}dr^2
+r^2(d\theta^2+\sin^2\theta d\phi^2)\  .
\end{equation}
In addition, according to the Birkhoff theorem, the curved
spacetime outside the spherically symmetric shell of radius $R$ is
described by the curved RN line element
\begin{equation}\label{Eq3}
ds^2=-\Big\{1-{{2[M+E(R)]}\over{r}}+{{Q^2}\over{r^2}}\Big\}dt^2+\Big\{1-{{2[M+E(R)]}\over{r}}+{{Q^2}\over{r^2}}\Big\}^{-1}dr^2
+r^2(d\theta^2+\sin^2\theta d\phi^2)\  .
\end{equation}

The (outer and inner) horizon radii of the central
charged black hole are characterized by the simple relation
\begin{equation}\label{Eq4}
r_{\pm}=M\pm (M^2-Q^2)^{1/2}\  .
\end{equation}
The Bekenstein-Hawking temperature of the black hole
is given by the compact expression \cite{Tem1,Tem2}
\begin{equation}\label{Eq5}
T_{\text{BH}}={{(r_+-r_-)\hbar}\over{2r^2_+}}\  .
\end{equation}

In the present paper we shall focus our
attention on extremal black holes.
These zero-temperature ($T_{\text{BH}}=0$)
charged black-hole configurations,
which mark the boundary between black-hole
spacetimes and horizonless naked singularities,
are characterized by the simple relation
\begin{equation}\label{Eq6}
Q=M\  .
\end{equation}

As explicitly shown in \cite{Hub}, the energy $E(R)$ of the massive shell
as measured by far away asymptotic observers can be deduced by solving the
equation of motion
\begin{equation}\label{Eq7}
\sqrt{g_{\text{in}}(r)+\dot R^2}-\sqrt{g_{\text{out}}(r)+\dot
R^2}=-{{m}\over{r}}\
\end{equation}
of the shell in the curved black-hole spacetime, where [see Eqs.
(\ref{Eq2}) and (\ref{Eq3}) with $Q=M$]
\begin{equation}\label{Eq8}
g_{\text{in}}(r)=\Big(1-{{M}\over{r}}\Big)^2\  ,
\end{equation}
\begin{equation}\label{Eq9}
g_{\text{out}}(r)=1-{{2[M+E(R)]}\over{r}}+{{M^2}\over{r^2}}\  ,
\end{equation}
and $\dot R\equiv dR/d\tau$ \cite{Notetau}.

\section{The minimal mass gap ${\cal E}_{\text{gap}}\equiv M-Q$ of near-extremal black holes}

In the present section we shall use analytical techniques in order
to provide supporting evidence for the existence of a finite mass gap,
\begin{equation}\label{Eq10}
{\cal E}_{\text{gap}}\equiv M-Q\  ,
\end{equation}
for a near-extremal RN black hole above the extremal limit
(\ref{Eq6}). To this end, we shall analyze a gedanken experiment in
which a shell of proper mass $m$, which is concentric with an
extremal charged RN black hole, is lowered towards the central black
hole. At some point, the shell would merge with the original
extremal black hole to form a larger black hole [see Eq.
(\ref{Eq14}) below], thus increasing the mass (energy) of the central black hole. Below we
shall try to {\it minimize} the energy increase (\ref{Eq10}).

In order to determine the minimal mass gap of near-extremal black
holes above the extremal limit (\ref{Eq6}), one should try to
deliver to the central extremal black hole the smallest possible
(non-zero) amount of energy. We shall therefore lower the massive
shell slowly towards the central black hole with an infinitesimally
small radial velocity
\begin{equation}\label{Eq11}
\dot R\to0\  .
\end{equation}
As we shall explicitly show below, most (but {\it not}
all) of the mass-energy of the shell as measured by asymptotic
observers would be red-shifted by the gravitational field of the
central black hole during this adiabatic lowering process. 
It is important to emphasize that the physical apparatus which is required in order 
to perform the adiabatic lowering process of the shell towards the black hole 
would probably need to withstand extremely high tensions \cite{Bekchal}. Thus, it should be realized that the present gedanken 
experiment is a highly idealized one and technical problems may prevent the actual performance of the experiment. 

Taking cognizance of Eqs. (\ref{Eq7}), (\ref{Eq8}), (\ref{Eq9}), and
(\ref{Eq11}), one finds that the total energy (as measured by far
away asymptotic observers) of the massive shell in the curved
black-hole spacetime is given by the simple functional expression
\begin{equation}\label{Eq12}
E(R)=m\cdot\Big(1-{{M}\over{R}}\Big)-{{m^2}\over{2R}}\ ,
\end{equation}
where $R$ is the radius of the shell. The first term on the r.h.s of
(\ref{Eq12}) represents the red-shifted mass-energy of the shell in
the curved black-hole spacetime. The second term on the r.h.s of
(\ref{Eq12}) represents the gravitational self-energy (an energy
term proportional to $m^2$) of the massive shell.

It is worth noting that {\it if} the massive shell could be lowered
slowly all the way down to the horizon ($r_{\text{H}}=M$) of the
central extremal RN black hole, then the mass term $m\cdot(1-{{M}/{R}})$ in (\ref{Eq12}) would be
completely red-shifted in this process [see, in particular, the energy expression
(\ref{Eq12}) with $R=r_{\text{H}}=M$]. However, as we shall now
prove explicitly, the adiabatic lowering process of the massive
shell towards the central black hole must stop in the regime $R>M$
{\it before} the shell reaches the original horizon $r_{\text{H}}=M$
of the central extremal black hole.

In particular, it is easy to verify that the lowering process of the
shell towards the central black hole produces a new (and larger)
horizon, which is characterized by the relation $r_{\text{NH}}>M$,
{\it before} the massive shell reaches the radius $r_{\text{H}}=M$
of the original black-hole horizon. Taking cognizance of the curved
line element (\ref{Eq3}), one finds that the physical condition for
the formation of a new engulfing horizon in the composed
black-hole-massive-shell system is given by the characteristic
functional expression
\begin{equation}\label{Eq13}
1-{{2[M+E(R)]}\over{R}}+{{Q^2}\over{R^2}}=0\ \ \ \ \text{with}\ \ \
\ Q=M\  .
\end{equation}
Substituting the $R$-dependent energy (\ref{Eq12}) of the shell into
the characteristic black-hole relation (\ref{Eq13}), one deduces
that a new and larger horizon, which engulfs both the original
central black hole and the massive shell, is formed in the composed
spherically symmetric black-hole-massive-shell system when the
radius of the shell reaches the critical value
\begin{equation}\label{Eq14}
R\to r_{\text{NH}}\equiv M+m\  .
\end{equation}
The radius (\ref{Eq14}) of the new horizon in the composed
black-hole-shell system is larger than the original radius
$r_{\text{H}}=M$ of the central extremal black hole.

Before proceeding, it is worth stressing the fact that the
analytically derived expression (\ref{Eq14}) provides the minimum
possible radius of the new engulfing horizon that results when a
shell of mass $m$ is added to the central extremal RN black hole. In
particular, the total energy of a massive shell which has a
non-vanishing kinetic energy (a non-vanishing radial velocity) is
larger than the one given by the energy expression (\ref{Eq12}).
Hence, a massive shell which falls towards the central extremal
black hole with a non-vanishing radial velocity would form an
engulfing new horizon which is {\it larger} than the one given by
the relation (\ref{Eq14}) \cite{Notelb}.

Substituting Eq. (\ref{Eq14}) into Eq. (\ref{Eq12}) one finds that,
for a given mass $M$ of the original central black hole, the energy
gap ${\cal E}_{\text{gap}}$ (the minimum energy which is delivered
to the extremal black hole due to the assimilation of the massive
shell) is given by the remarkably simple functional expression
\begin{equation}\label{Eq15}
{\cal E}_{\text{gap}}=E_{\text{min}}(m;M)={{m^2}\over{2(M+m)}}\ .
\end{equation}
The energy expression (\ref{Eq15}) implies that the minimum energy $E_{\text{min}}(m;M)$ which is
delivered to the extremal RN black hole in the present gedanken experiment can be minimized by
minimizing the mass $m$ of the shell. We therefore raise here the
following physically important question: How {\it small} can the
mass $m$ of the shell be?

In order to address this physically interesting question, we can
think of the shell as representing a spherically symmetric localized
wave packet of a massive particle. It is well known that, according
to quantum theory and special relativity, a wave packet representing
a particle of mass $m$ cannot be localized to better than its
Compton length: $R\geq \hbar/2m$ \cite{Com1,Com2}. This relation
implies that the circumference $C=2\pi R$ of the spherically
symmetric wave packet is bounded from below by the relation
\begin{equation}\label{Eq16}
2\pi R \geq {{h}\over{2m}}\  .
\end{equation}
One therefore concludes that a spherically symmetric
compact wave packet of circumference radius $R$ representing a
particle of mass $m$
is characterized by the lower bound
\begin{equation}\label{Eq17}
m\geq {{\hbar}\over{2R}}\  .
\end{equation}

Since in our case the radius of the shell at the point of
assimilation is $R_{\text{min}}=M+m$ [see Eq. (\ref{Eq14})], one
finds the simple expression \cite{Noteap1}
\begin{equation}\label{Eq18}
m_{\text{min}}={{\hbar}\over{2M}}\cdot[1+O(\hbar/M^2)]\
\end{equation}
for the minimally allowed mass of the shell which is consistent with
quantum theory and special relativity. Substituting (\ref{Eq18})
into (\ref{Eq15}), one finds the expression
\begin{equation}\label{Eq19}
{\cal E}_{\text{gap}}=E_{\text{min}}={{\hbar^2}\over{8M^3}}\
\end{equation}
for the energy gap of the central extremal black hole in the regime
$M^2\gg\hbar$ of large black holes. 
It is important to emphasize that we cannot rule out the 
possible existence of a different physical process that may yield an energy gap which 
may be smaller than the analytically derived energy gap (\ref{Eq19}). 

\section{Summary}

Using a sophisticated string theory analysis, Maldacena and Susskind
\cite{Gap1} have nicely shown that near-extremal black holes are
characterized by a {\it finite} mass gap of the first excited state
above the extremal (zero-temperature) black-hole configuration. 
As explicitly proved in the physically interesting work \cite{Gap2}, the same energy gap can be inferred 
using a physical argument which is based on holography 
without using explicit string theory calculations. 

The main goal of the present compact paper was to gain some physical
insights on the origin of the intriguing mass gap (\ref{Eq1}). In
particular, we have demonstrated that the suggested 
finite mass gap (\ref{Eq1}) can be deduced from a simple
{\it semi-classical} gedanken experiment in which a spherically
symmetric massive shell is lowered towards a central extremal
Reissner-Nordstr\"om black hole which eventually absorbs the shell.
In order to determine the {\it minimum} mass gap of the system above
the extremal black-hole limit (\ref{Eq6}), we have tried to minimize
the energy which is delivered to the black hole by the shell. We
have therefore assumed that the massive shell is lowered
adiabatically (slowly) towards the central black hole.

The main results derived in the present paper and their physical
implications are as follows:
\newline
(1) It has been shown that during the lowering process of the shell,
most (but {\it not} all) of its mass-energy as measured by far away
asymptotic observers is red-shifted by the gravitational field of
the central black hole. Intriguingly, however, we have explicitly
proved that at some point during the lowering process of the shell,
a new and larger horizon, which engulfs {\it both} the original
black hole and the massive shell, is formed. In particular, the
engulfing new horizon with $r_{\text{NH}}>M$ [see Eq. (\ref{Eq14})]
always forms {\it before} the massive shell reaches the horizon
$r_{\text{H}}=M$ of the original extremal black hole.
\newline
(2) The formation of the new and larger horizon prevents one from
totally red-shifting the mass-energy of the lowered shell. Thus, for
a given proper mass $m$ of the shell, there is always a finite ({\it
non}-zero) lower bound $E_{\text{min}}(m;M)=m^2/[2(M+m)]$ [see Eq. (\ref{Eq15})] on the
energy which is delivered to the central extremal black hole.
\newline
(3) Taking cognizance of the fact that a combination of quantum
theory and special relativity sets the lower bound $R \geq \hbar/2m$
on the proper circumference radius of a shell of proper mass $m$, we
have derived the semi-classical expression 
\begin{equation}\label{Eq20}
{\cal E}_{\text{gap}}=M-Q={{\hbar^2}\over{8M^3}}\
\end{equation}
for the minimum energy which is delivered to the central extremal black hole in the present gedanken experiment.

Interestingly, the results of the present semi-classical gedanken experiment support the validity of the 
previously derived \cite{Gap1,Gap2} quantum energy gap (\ref{Eq1}) 
of extremal black holes in a wider context than string theory and holography. As emphasized above, 
our analysis cannot rule out the 
possible existence of another physical process that may yield an energy gap for extremal black holes 
which may be smaller than the characteristic gap (\ref{Eq20}). 

\bigskip
\noindent
{\bf ACKNOWLEDGMENTS}
\bigskip

This research is supported by the Carmel Science Foundation. I thank
Yael Oren, Arbel M. Ongo, Ayelet B. Lata, and Alona B. Tea for
stimulating discussions.


\begin{thebibliography}{99}
\bibitem{Noteunits} We use natural units in which $G=c=1$. In these
units, mass, horizon radius, electric charge, and specific angular
momentum all have the same units.
\bibitem{Noteq0} We shall assume $Q>0$ for the black-hole electric charge without loss of
generality.
\bibitem{Notemq} Note that a zero temperature (extremal) RN black
hole is characterized by the simple relation $Q=M$ [see Eq.
(\ref{Eq6}) below].
\bibitem{Notesch} Here $\{t,r,\theta,\phi\}$ are the standard Schwarzschild coordinates.
\bibitem{Notetau} Here $\tau$ is an affine parameter along the worldline of the shell \cite{Hub}.
\bibitem{Notelb} For example, a spherically symmetric shell of mass $m$
which starts falling from rest at spatial infinity towards the
central extremal black hole would form a new engulfing horizon when
its radius shrinks to $r_{\text{NH}}=M+m+[(M+m)^2-M^2]^{1/2}$. This
horizon radius is larger than the one given by Eq. (\ref{Eq14}).
\bibitem{Noteap1} Note that from Eqs. (\ref{Eq14}) and (\ref{Eq17}) one
finds the lower bound $m\geq
m_{\text{min}}=(\sqrt{4M^2+8\hbar}-2M)/4$ on the mass of the shell
in our gedanken experiment, which yields the simple relation
(\ref{Eq18}) in the regime $M^2\gg\hbar$ of large black holes.
\end{thebibliography}
\end{document}